\documentclass[twoside,nohyper]{JHEP3}
\usepackage{amsmath,amssymb,longtable,dsfont}

\newcommand{\sh}{\not\!\!}
\newcommand{\lag}{{\mathcal L}}
\newcommand{\ve}{\varepsilon}
\newcommand{\ut}{\pmb{1}}

\newcommand{\beq}{\begin{equation}}
\newcommand{\eeq}{\end{equation}}
\newcommand{\bea}{\begin{eqnarray}}
\newcommand{\eea}{\end{eqnarray}}
\newcommand{\bes}{\begin{split}}
\newcommand{\ees}{\end{split}}
\newcommand{\bed}{\begin{displaymath}}
\newcommand{\eed}{\end{displaymath}}
\newcommand{\od}{{\cal O}(q^2)}
\newcommand{\ot}{{\cal O}(q^3)}
\newcommand{\cd}{\chi_{\rm dim}}
\newcommand{\mb}{{\mathcal L}_{MB}}
\newcommand{\la}{\langle}
\newcommand{\ra}{\rangle}
\newcommand{\nn}{\nonumber}

\newcommand{\al}{\alpha}
\newcommand{\be}{\beta}
\newcommand{\ga}{\gamma}
\newcommand{\rh}{\rho}
\newcommand{\et}{\eta}
\newcommand{\si}{\sigma}

\title{Meson-Baryon Effective Chiral Lagrangians to ${\cal O}(q^3)$}
\author{Jos\'e Antonio Oller and Michela Verbeni\\
Departamento de F\'\i sica, 
Universidad de Murcia, E-30071 Murcia, Spain\\
E-mail: \email{oller@um.es} , \email{mverbeni@ugr.es}}
\author{Joaquim Prades\\
CAFPE and Departamento de F\'\i sica Te\'orica y del Cosmos,
Universidad de Granada,\\
Campus de Fuente Nueva, E-18002 Granada, Spain\\
E-mail: \email{Prades@ugr.es}}

\abstract{We construct the complete and minimal ${\cal O}(q^2)$ and ${\cal O}(q^3)$  three-flavour 
Lorentz invariant chiral
 meson-baryon 
Lagrangians for the first time in the literature. We compare with previous three-flavour studies 
 reducing the number of independent monomials and adding new ones that
 were missing.}

\keywords{Chiral Lagrangians, NLO Calculations, QCD}
\preprint{CAFPE/69-06\\ UGFT/199-06\\July 2006}

\numberwithin{equation}{section}
\begin{document}

\section{Introduction}
The extension of Chiral Perturbation Theory (CHPT) \cite{fisica,glsu2,gl}
 to the one baryon sector is not straightforward, since, employing 
dimensional regularization, higher order loops contribute 
 to lower order calculations. This is a consequence of the fact that
 the nucleon mass does
 not vanish in the chiral   limit \cite{gss}, therefore 
the correspondence between loop and  
 chiral expansion is lost. This shortcoming was overcome within the 
formalism of    heavy baryon CHPT \cite{anis,ulf}, where most 
of the higher order calculations  in baryon CHPT have been performed 
(for reviews, see e.g. \cite{6,7}). More recently, it was realized 
that  chiral power counting and loop expansion can be reconciled with a 
Lorentz invariant formulation  of baryon CHPT employing the so called infrared 
regularization \cite{ir,ellis}.
In the literature have appeared many one loop calculations realized
employing this scheme, especially in $SU(2)$
baryon CHPT \cite{17,22,27,29,30}. This method has also 
the advantage of correctly keeping the analytical properties 
of physical amplitudes, that in some cases are lost in heavy baryon 
CHPT in the low energy region. On the other hand, the chiral 
pion-nucleon $SU(2)$ Lagrangian is completely  known up to ${\cal O}(q^4)$ 
\cite{fettes00}, both the relativistic and the heavy baryon
projected.

In baryon CHPT, the two flavour effective field theory is more developed
than the three flavour one. Actually in this case 
the  relative large kaon mass makes no  clear a 
priori whether the meson-baryon system 
can be treated  perturbatively. Furthermore
in this sector one has also
to face the presence  of resonances  close to or even below 
the pertinent thresholds, 
e.g. the $\Lambda(1405)$.\footnote{The implementation
 of non-perturbative resummation methods  within 
the chiral expansion  has allowed    the successful use  of chiral 
Lagrangians for the  study of scattering and production processes 
in  $SU(3)$ baryon CHPT
\cite{kaiserw,npa,oset,oller,borasoy,comment,nostroprl,reply,
Oller06,BMN06,jnp}.} Most of the calculations in  $SU(3)$
baryon CHPT have been performed within the heavy baryon approximation
 \cite{borasoyann,frink,mmstei,fmku,kpstei,sku}. In  \cite{mueller} 
the complete renormalization   of the generating functional 
for Green functions of quark currents  between one baryon states 
in three flavour heavy baryon CHPT is performed  up to $\ot$. 
Some calculations have been already done within   the infrared regularization 
scheme  \cite{frinkmas,ellis,fmirku,fmirkub,afmpug,afmpug2} or within 
the extended on-mass-shell renormalization scheme
\cite{schindler,sch2}.
 
An important aspect of this relative lack of development of $SU(3)$
baryon CHPT is
the unsatisfactory way 
the $\od$ and, particularly, 
the $\ot$ meson-baryon Lorentz invariant  chiral 
Lagrangians are given in the literature. 
The main purpose of this work consists in filling this gap.
Since its publication, 
Krause's work  \cite{krause} has   been employed as a standard 
reference for the effective Lorentz invariant chiral meson-baryon
  Lagrangian  with three flavours up to $\ot$. However, 
the number of monomials appearing there can be further reduced, 
as shown below in  section \ref{lagrangians}. 
  Furthermore,  the presentation of the monomials given 
in \cite{krause} can be certainly improved  allowing for a much
  easier manipulation. At $\od$ part of the meson-baryon effective 
chiral Lagrangian 
is given without derivation in  \cite{frinkmas}. Again, 
 we find that this Lagrangian    can be further reduced 
and given in more compact  form.

The content of the paper is organized as follows. 
In section \ref{sec:2} we present 
the building blocks that will be used 
in the construction of the effective meson-baryon Lagrangian and then we
discuss their symmetry properties in 
section \ref{interaction_terms}. In this section we also establish 
 the conditions  to be obeyed by 
the monomials written with the building blocks, in order to obtain
 a Lagrangian invariant under the 
 strong interaction symmetries. All the general relations 
 employed to reduce the  number  of independent monomials 
  are listed in section \ref{construction}.
 More specific manipulations are given in appendix \ref{elimination}. 
Our final expressions for the $\od$  and $\ot$ Lagrangians are displayed  
in section \ref{lagrangians}.
 Finally, in section \ref{conc} we summarize our main conclusions. 
 
\section{General Framework and Building Blocks}
\label{sec:2}
The procedure for constructing non-linear 
effective chiral symmetric Lagrangians is standard \cite{CWZ69}.
We briefly sketch this procedure below.

QCD with three massless quarks, $u$, $d$, and $s$, 
exhibits a global $SU(3)_L\otimes SU(3)_R$ chiral symmetry, 
which is spontaneously broken to the subgroup $SU(3)_V$, with $V=L+R$. 
In order to write down the chiral invariant effective 
Lagrangian, it is convenient to promote the chiral symmetry 
to a local one introducing  external hermitian $3\times 3$ 
 matrix fields $s(x)$, $p(x)$, $v_\mu(x)$ and $a_\mu(x)$ 
which  couple  to  scalar,  pseudoscalar, vector and axial  
quark currents, respectively,  as follows
\beq
\lag = \lag^0_{\text{QCD}} + \bar{q}\gamma^\mu (v_\mu + 
\gamma_5 a_\mu )q - \bar{q}(s-i \gamma_5 p) q.
\label{lagqcd}
\eeq
Here, $\lag^0_{\text{QCD}}$ is the QCD Lagrangian with 
massless $u$, $d$ and  $s$ quarks and 
 current quark masses appear in the scalar source as  
$s(x)= {\cal M}+\cdots$, where ${\cal M}=\hbox{\rm diag}(m_u,m_d,m_s)$
is a $3\times 3$ matrix collecting the light quark masses.
 For the construction of the  $SU(3)_L \otimes SU(3)_R$ chiral
invariant Lagrangian we impose  the constraints 
$\la a_\mu\ra=\la v_\mu\ra=0$.\footnote{Here and in the rest of the
paper,  $\la X \ra$  stands for the flavour trace of $X$.}
 Electromagnetic interactions are introduced through 
the external vector field $v_\mu=|e|Q A_\mu$,
 where  $Q={\rm diag}(2,-1,-1)/3$ is the quark
electrical charge matrix and $A_\mu$ the photon
 field -- notice that $\la v_\mu\ra =0$ in this case.

The  $SU(3)$ effective chiral Lagrangian 
describing the interactions of  the lightest pseudoscalar  meson
and baryon octets and external  sources (photons, $\dots$) is 
obtained by constructing the most general  Lagrangian
which is invariant under $SU(3)_L \otimes SU(3)_R$ transformations and 
satisfies strong interaction symmetries.

The relevant degrees of freedom in  the 
effective meson-baryon Lagrangian 
are  the spontaneous chiral symmetry breaking  Goldstone bosons  
and the octet of $J^P =  \frac{1}{2}^+$ baryons.  Goldstone bosons 
are represented by a matrix field $u(\Phi)$ which transforms
under a general chiral rotation $g=(g_L,g_R)\in SU(3)_L\otimes
SU(3)_R$ as
\beq 
u\longrightarrow u^\prime = g_R\, u \,h^{\dagger}(g,u) = h(g,u)\,
 u\, g^{\dagger}_L
\label{utrans}
\eeq
 according to the standard non-linear realization \cite{CWZ69}, with  
 $h(g,u)\in SU(3)_V$.
 We use the  standard parametrization for the  matrix field $u(\Phi)$, 
$u=\exp(i\Phi/\sqrt{2}F)$ with $\Phi$ given by,
\beq
\Phi=\left(
\begin{array}{ccc}
\frac{\pi^0}{\sqrt{2}}+\frac{\eta_8}{\sqrt{6}} & \pi^+ & K^+ \\
\pi^- & -\frac{\pi^0}{\sqrt{2}}+\frac{\eta_8}{\sqrt{6}} & K^0  \\
K^- & \overline{K}^0 & -\frac{2\eta_8}{\sqrt{6}}
\end{array}
\right)~.
\label{mesonoctet}
\eeq
The octet of $J^P = \frac{1}{2}^+$ baryons is arranged in a $3\times 3$
traceless matrix $B$, 
\beq
B=\left(
\begin{array}{ccc}
\frac{\Sigma^0}{\sqrt{2}}+\frac{\Lambda}{\sqrt{6}} & \Sigma^+ & p \\
\Sigma^- & -\frac{\Sigma^0}{\sqrt{2}}+\frac{\Lambda}{\sqrt{6}} & n  \\
\Xi^- & \Xi^0 & -\frac{2\Lambda}{\sqrt{6}}
\end{array}
\right)~
\label{baryonoctet}
\eeq
and corresponds to massive fields  in the adjoint $SU(3)_V$ representation 
transforming as
\beq
B\longrightarrow B'= h(g,u) B h^{\dagger}(g,u)
\label{Btrans}
\eeq
under chiral transformations \cite{CWZ69}.

The basic building blocks we use to construct the 
effective  chiral Lagrangian are 
\beq\begin{split}
u_\mu &= i\{u^\dagger (\partial_\mu-i
r_\mu)u-u(\partial_\mu-il_\mu u^\dagger) \}\, , \\
\chi_\pm &= u^\dagger \chi u^\dagger \pm u\chi^\dagger u, \\
f_\pm^{\mu\nu} &= u F^{\mu\nu}_L u^\dagger \pm u^\dagger F^{\mu\nu}_R
u~, \, 
\label{set}
\end{split}\eeq
where  $\chi = 2 B_0 \,(s+i p)$
and $B_0=-\la 0|\bar{q}q|0\ra/F^2$, with $\la 0|\bar{q}q|0\ra$ 
  the $SU(3)$ quark condensate and $F$ 
the pion weak decay constant, both in the chiral limit. 
Here,  
\begin{align}
F^{\mu\nu}_R &= \partial^\mu r^\nu - \partial^\nu r^\mu
-i[r^\mu,r^\nu] ,& r^\mu&=v^\mu+a^\mu~,\nn\\
F^{\mu\nu}_L &= \partial^\mu l^\nu - \partial^\nu l^\mu
-i[l^\mu,l^\nu] ,& l^\mu&=v^\mu-a^\mu~,
\end{align}
are the external field strength tensors. 
The matrices $u_\mu$, and $f_\pm^{\mu\nu}$ 
are traceless since  we impose $\la v_\mu\ra = \la a_\mu\ra =0$.  

The operators in (\ref{set}) or any product thereof 
transform under  $SU(3)_L \otimes SU(3)_R$ 
transformations  as $X\to h \, X \, h^\dagger$ and their covariant 
derivative reads  
\beq
D_\mu X =\partial_\mu X + [\Gamma_\mu, X],
\label{covder}
\eeq
where $\Gamma_\mu$ is the chiral connection,
\beq
\Gamma_\mu =
\frac{1}{2}\{u^\dagger(\partial_\mu-ir_\mu)u+u(\partial_\mu-il_\mu))
u^\dagger\}.
\eeq
We collectively call the operators
in \eqref{set} and their covariant derivatives ``chiral fields''.

For the construction of the effective Lagrangian the  two relations 
\bea
\label{curvature}
[D_\mu,D_\nu]  X &=&
\frac{1}{4}[[u_\mu,u_\nu], X ]-\frac{i}{2}[f^+_{\mu\nu}, X]~,\\
D_\nu u_\mu - D_\mu u_\nu~&=&f^-_{\mu\nu}~, 
\label{fmunuminus}
\eea
turn out to be very useful. 
 The  first relation allows to consider only symmetric products of 
covariant derivatives while the second one to take just 
symmetrized covariant derivatives acting on  $u_\mu$, 
\beq
h_{\mu\nu}=D_\mu u_\nu+D_\nu u_\mu~.
\eeq

\section{Construction of Allowed Monomials}
\label{interaction_terms}
The chiral dimension of the building blocks in \eqref{set} is
\beq\begin{split}
&~u_\mu  \sim {\cal O}(q)~,\\
& \chi_\pm,~f^\pm_{\mu\nu}\sim {\cal O}(q^2).  
\label{ch_dim}
\end{split}\eeq
The action of $n$ covariant derivatives  on any of the fields
in (\ref{set}) increases of $n$ units the chiral order.  We cannot
extend this chiral counting rule to the field $B$ as 
the covariant derivative,  when applied to a baryon field,
counts as a quantity of $O(q^0)$, since
 the baryon mass does not vanish in the chiral limit.
However, the combination $(i\sh\! D - M_0) B$, where $M_0$ is the
octet baryon mass in the chiral limit,  
can be considered a small quantity \cite{krause} of the order of the 
soft momenta associated with pseudoscalar and external fields.
 Then we have the chiral counting rules
\beq\begin{split}
&B,~\bar{B},~ D_\mu B\sim {\cal O}(q^0),\\
&(i\sh D - M_0) B\sim {\cal O}(q)~.
\label{BChdim}
\end{split}\eeq
The elements of the Clifford algebra basis 
have the following chiral dimensions
\beq\begin{split}
&\ut,~\gamma_\mu,~\gamma_5\gamma_\mu,~\sigma_{\mu\nu} 
\sim {\cal O}(q^0)~,\\
& \gamma_5 \sim {\cal O}(q)~,
\label{chdim_clif}
\end{split}\eeq
as $\gamma_5$ couples the small and the large components of 
the baryon spinor. We refer to
 the assignment of chiral dimensions in 
the baryonic sector given in eqs. \eqref{BChdim} and 
\eqref{chdim_clif} as  the covariant chiral counting.

The transformation properties under parity (P), charge conjugation (C) 
and hermitic conjugation (h.c.)  of the building blocks 
in \eqref{set} can be found in table \ref{prop_mes}, while in table 
\ref{prop_gamma},  we give the corresponding properties of the 
matrices $\Gamma$ in (\ref{chdim_clif}) when appearing in the 
 baryon bilinear $\la \bar{B}\Gamma B\ra$.
\TABLE{
\begin{tabular}{|c|c|c|c|c||c|c|c|}
\hline
& P& C&h.c.& $\cd$ & $p$ & $c$ & $h$\\
\hline
 $u_\mu$ & $-P_\mu^\nu u_\nu$ &$u_\mu^T$  & $u_\mu$& 1& 1 & 0 & 0\\
  \hline
$f^+_{\mu\nu}$ & $ P_\mu^\lambda P_\nu^\sigma f^+_{\lambda\sigma}$  
& $- f^{+T}_{\mu\nu}$ & $f^+_{\mu\nu}$&2& 0 & 1 & 0\\
 \hline
 $f^-_{\mu\nu}$ & $-P_\mu^\lambda P_\nu^\sigma f^-_{\lambda\sigma}$ 
& $f^{-T}_{\mu\nu}$& $f^-_{\mu\nu}$&2& 1 & 0 & 0\\
\hline
$\chi_+$&$\chi_+$ & $\chi_+^T$&$\chi_+$&2& 0 & 0 & 0\\
\hline
$\chi_-$&$- \chi_-$&$\chi_-^T$&$- \chi_-$&2 & 1 & 0 & 1\\
\hline
$\overrightarrow{D}_\mu$ & $P_\mu^\nu \overrightarrow{D}_\nu$& 
$\overleftarrow{D}_\mu^T$&$\overleftarrow{D}_\mu$&1& 0 & 0 & 0\\
\hline
\end{tabular}
\caption{Parity (P), charge conjugation (C) and 
hermitic conjugation (h.c.)  transformation properties 
 and chiral dimension of the building blocks 
 and of their covariant derivative. 
$P_\mu^\nu\equiv {\rm diag} (+1,-1,-1,-1)$ is 
the matrix  associated with the parity operator. 
See \protect{\eqref{P}}, 
\protect{\eqref{CC_WO_der}} and \protect{\eqref{HC_WO_der}} 
for the definition of $p$, $c$ and $h$.
  \label{prop_mes}}}
\TABLE{
\begin{tabular}{|c|c||c|c|c|}
\hline
&$\cd$ & $p$ & $c$ &$h$\\
\hline
$\ut$ & 0 & 0 & 0 & 0\\
\hline
$\gamma_5$ & 1 &  1 & 0 & 1\\
\hline
$\gamma_\mu$ & 0 & 0 & 1 & 0 \\
\hline
$\gamma_5\gamma_\mu$ & 0 & 1 &0 & 0\\
\hline
$\sigma_{\mu\nu}$ &0 & 0 & 1 & 0\\
\hline
\end{tabular}
\caption{Parity (P), charge conjugation (C) and 
hermitic conjugation (h.c.) transformation properties 
and chiral dimension  of the Clifford algebra elements. 
See \protect{\eqref{P}}, 
\protect{\eqref{CC_WO_der}} and \protect{\eqref{HC_WO_der}} 
for the definition of $p$, $c$ and $h$.
\label{prop_gamma}}}

We start by  writing down all possible chiral symmetric monomials
fulfilling strong interaction symmetries, that is which are
invariant under Lorentz and parity transformations and charge and 
hermitic conjugation.
  A generic term is a bilinear in baryon fields and can contain 
more than  one trace in flavour space.
For every term in the Lagrangian being a Lorentz scalar, 
the space-time indices coming from chiral fields, covariant derivatives, 
Clifford algebra basis elements and  tensors  $g_{\mu\nu}$ 
 and/or pseudotensors $\ve_{\mu\nu\alpha\beta}$, 
must be suitably contracted.

We first consider monomials composed by one trace and afterwards 
we discuss the case with two traces. Since matrix fields do not
commute, we have to take into account all possible orderings.
 To this end and to have terms  whose transformation properties under 
charge and hermitic conjugation are easily obtained, 
it is convenient to employ the  form:
\begin{equation}
X=\la \bar{B} (A_1,\dots, (A_n,\Theta  D^mB)\dots)\ra.
\label{1trgen}
\end{equation}
The fields $A_1,A_2,\dots,A_n$ can be single chiral fields or a
combination of (anti)commutators
 thereof and $(A_i,A_j)$ denotes either the commutator, $[A_i,A_j]$, 
or the anticommutator, $\{A_i,A_j\}$, of $A_i$ and $A_j$.  
 The symbol $\Theta$ indicates the product of  an element
 of the Clifford algebra basis, $\Gamma$, 
times metric tensors  and/or  Levi-Civita pseudotensors
while $D^m$  is a set of $m\geq 0$ covariant 
derivatives acting on $B$ in a totally  symmetrized way. In the previous
 equation, for the sake of simplicity in the notation,
 we have not shown explicitly the space-time indices attached to 
$A_i$, $\Theta$ and $D^m$.

The invariance of a candidate monomial $X$ under P is easily checked
taking into account the following transformation  properties under parity 
\beq\begin{split}
&\la \bar{B} (A_1,\dots,
(A_n,\Theta D^m
B)\dots)\ra^P\\
&\quad =(-1)^{p_1+\dots+p_n+p_\Gamma+n_\ve}
\la \bar{B} (A_1,\dots, (A_n,\Theta D^m B)\dots)\ra,
\label{P}
\end{split}\eeq
where $n_\ve$ is the number of Levi-Civita pseudotensors
present in \eqref{1trgen} and the values of the exponents follow from
tables \ref{prop_mes} and \ref{prop_gamma}. 
The subscript in $p_\Gamma$
refers to the   Clifford algebra matrix $\Gamma$  contained 
in $\Theta$, as explained above. From \eqref{P}, it follows
that a candidate term can occur in $\lag_{MB}$ only if
\beq
(-1)^{p_1+\dots+p_n+p_\Gamma+n_\ve}=1~.
\label{cond_P}
\eeq

We next examine how $X$ transforms under charge and hermitic
conjugation. Here we essentially follow the lines of the analysis  
in ref.~\cite{krause}. We first consider the case without covariant
derivatives acting on the baryon fields. Under
charge conjugation the monomial \eqref{1trgen} transforms as
\beq\begin{split}
 &\la \bar{B} (A_1,\dots,
 (A_n,\Theta B)\dots)\ra^C \\
 &\quad= (-1)^{c_1+\dots c_n+c_\Gamma}\la \bar{B} (A_n,\dots,
(A_1,\Theta B)\dots)\ra,
\label{CC_WO_der}
\end{split}\eeq
where $c_i$ and $c_\Gamma$ are determined from tables 
\ref{prop_mes} and \ref{prop_gamma},  respectively. 
Analogously, under hermitic conjugation, we have 
\beq\begin{split}
 &\la \bar{B} (A_1,\dots,
 (A_n,\Theta B)\dots)\ra^\dagger \\
 &\quad= (-1)^{h_1+\dots h_n+h_\Gamma}\la \bar{B} (A_n,\dots,
(A_1,\Theta B)\dots)\ra,
\label{HC_WO_der}
\end{split}\eeq
with $h_{i}$ and $h_\Gamma$ determined again from tables 
\ref{prop_mes} and \ref{prop_gamma}, respectively.
 Using the identities
\beq\begin{split}
[A,[C,B]]&=[C,[A,B]]+[[A,C],B]\\
[A,\{C,B\}]&=\{C,[A,B]\}+\{[A,C],B\}\\
\{A,\{C,B\}\}&=\{C,\{A,B\}\}+[[A,C],B],
\label{tr_id}
\end{split}\eeq
we can bring the terms in the r.h.s. of
eqs. \eqref{CC_WO_der} and \eqref{HC_WO_der} to a form 
in which the operators $A_i$ appear in the same order as in the original
monomial, plus additional pieces:
\beq\begin{split}
&\la\bar{B}(A_n,\dots,(A_2,(A_1,\Gamma B)
  \dots))\ra\\
&\quad=\la\bar{B}(A_1,(A_2,\dots,(A_n,\Gamma  B)\dots))\ra
+\la\bar{B}(\tilde{A}_1,\dots,(\tilde{A}_2,(\tilde{A}_m,\Gamma  B))
  \dots )\ra,
\label{reorder}
\end{split}\eeq
where $\tilde{A}_i$ are (anti)commutators of the
fields $A_i$, with $m<n$. 
 To guarantee charge conjugation 
invariance of the effective interaction 
constructed from the monomial $X$, the combination 
$(X+X^C)/2$ must be taken.
 From this consideration and  eqs. 
\eqref{CC_WO_der} and \eqref{reorder},
we conclude that a term $X$ as defined in (\ref{1trgen})
will appear in $\mb$ only if
\beq
(-1)^{c_1+\dots+c_n+c_\Gamma}=1.
\label{cond_CC_WO_der}
\eeq

Using  (\ref{CC_WO_der}) and (\ref{HC_WO_der}), it is easy to show
that  charge conjugation symmetric terms  are  either hermitian 
or anti-hermitian.  

We now consider the possibility that type \eqref{1trgen} monomials  
contain  $m$ covariant derivatives
acting on the baryon field $B$. In this case
under charge conjugation the monomial $X$ transforms as
\beq
X^C=(-1)^{c_1+\dots+c_n+c_\Gamma} \la
 \bar{B} \overleftarrow{D}^m (A_n,\dots,(A_1,\Gamma
 B)) \ra.
 \label{ccexact}
\eeq
After performing an integration by parts and eliminating a total
derivative, we can apply Leibniz rule and obtain 
a term with the covariant derivatives acting again on $B$ 
together with a sum of terms in which additional covariant derivatives 
operate on the chiral fields. According to the
chiral counting in the mesonic sector, the latter are, at least,
 of one order higher, so that we end up with
\beq
X^C=(-1)^{c_1+\dots+c_n+c_\Gamma+m} \la
 \bar{B} (A_n,\dots,(A_1,\Theta D^m B)) \ra + \text{h.o.}~,
\label{CC_der}
\eeq
where h.o. denotes higher order terms with covariant 
derivatives 
acting on the chiral fields $A_i$. 
Up to the order considered,  
these higher orders contributions can be neglected and the monomial $X$
will appear in $\mb$ only if 
\beq
(-1)^{c_1+\dots+c_n+c_\Gamma+m}=1~.
\label{cond_CC_der}
\eeq
This condition also explains why the covariant derivative acting on the
baryon field is considered odd under charge and hermitic conjugation.
However, the resulting effective interaction $(X+X^C)/2$,
where $X^C$ is given in (\ref{CC_der}) 
by removing the higher order terms, is not always exactly  
invariant under charge conjugation, but 
only up to the considered chiral order. Importantly, in the effective 
meson-baryon  chiral Lagrangian we choose 
to have terms that are exactly invariant under charge conjugation, i.e., 
to keep also the higher order contributions in (\ref{CC_der}), thus
the exact $X^C$ as given 
in eq. (\ref{ccexact}) is used.  
In this way,   the amplitudes calculated with $\mb$ 
will obey exact  crossing symmetry under the exchange of meson fields.
 This is, of course, a fundamental
property of physical amplitudes and is well worth keeping it exactly. 

In the effective Lagrangian, there can also appear terms 
which are the  products  of two or more flavour traces.
Explicitly, they can be either the product of one term of type
\eqref{1trgen} times flavour traces of chiral fields 
or  monomials  where  the $\bar{B}$ and $B$ matrix fields  are
contained in two different flavour traces.
Thus a general monomial can have one of the
following forms:
\bea
X_1&=&\la\bar{B} (A_1,\dots,(A_j,\Theta D^mB)\dots )\ra
\la (A_{j+1},\dots,(A_{n-2},A_{n-1})\dots)A_n\ra\label{2tr_1} \, ; \\
X_2&=&\la\bar{B}(A_1,\dots,(A_{j-1},A_j)\dots)\ra
\la(A_{j+1},\dots,(A_{k-1},A_k)\dots)\Theta D^m B\ra\nn\\
&&\quad\times\la (A_{k+1},\dots,(A_{n-2},A_{n-1})\dots)A_n\ra.\label{2tr_2}
\eea
One can have  more traces involving chiral fields than those 
explicitly shown above; in these cases the extension of the discussion 
below is  straightforward.

For $X_1$-type terms,  parity transformation, charge and hermitic conjugation 
properties can be studied analogously to the one flavour trace case
 and conditions (\ref{cond_P}) 
and (\ref{cond_CC_der}) must be satisfied  for these terms too. 
  For $X_2$-type terms, i.e., with $B$ and $\bar{B}$ in different traces,
one obtains that condition (\ref{cond_P}) has  to be satisfied
for  transformations under parity but
condition  (\ref{cond_CC_der})  for transformations under 
charge conjugation changes.
This is due to the fact that under  charge conjugation  
the monomial transforms as
\begin{eqnarray}
X^C_2&=&(-1)^{c_1+\dots+ c_n+c_\Gamma+m}
\la\bar{B}(A_{j+1},\dots,(A_{n-1},A_n)\dots)\ra
\la (A_1,\dots,(A_{j-1},A_j)\dots)\Theta D^mB\ra\nonumber\\
&\times&\la (A_{k+1},\dots,(A_{n-2},A_{n-1})\dots)A_n\ra+\hbox{h.o.}
\label{Ccon_2tr}
\end{eqnarray}
The monomial $X_2$ can always appear in $\mb$, 
even if condition \eqref{cond_CC_der} is not satisfied
 since  it is not possible,  using the (anti)commutator 
identities (\ref{tr_id}), 
 to reobtain the original term. As in the case with only one 
 trace, we will take the combination $(X_i+X_i^C)/2$ with 
exact $X_i^C$, $i=1$, 2.    As in that
case and both for $X_1$ and $X_2$,  it is easy to show
that charge conjugation invariant terms are either 
hermitian or anti-hermitian. 

\section{Construction of the Effective Chiral Meson-Baryon
  Lagrangian}\label{construction}
In this section, we outline the method employed 
to get a minimal set of effective  meson-baryon
monomials up to $\ot$. Listing the terms  satisfying the required
symmetry conditions is  a straightforward operation.
In $SU_L(3) \otimes SU_R(3)$, at this order, with $\la
a_\mu\ra=\la v_\mu\ra=0$, 
we just need to consider monomials with one and two flavour traces.  
The procedure we use to obtain a complete list of allowed monomials
is as follows. For a fixed element  of the Clifford  
algebra basis (\ref{chdim_clif}) and number of flavour traces,
we write down {\em all} possible monomials with  
the smallest number of covariant derivatives acting on the 
baryon field $B$ that fulfill  
the symmetry requirements discussed in the previous section. 
The number of covariant derivatives acting on $B$ is then gradually increased 
 for the same  Clifford  algebra basis  element and number 
of flavour traces. The procedure
is over when the addition of more covariant derivatives acting on $B$
does not yield
 new independent monomials due to the relations 
(\ref{first_rel})-(\ref{seventh_rel}) given below.

Once a complete list of allowed monomials 
is obtained, the main task consists in    
finding  out a minimal set of linearly independent interaction terms.
In order to minimize the number of terms, we extensively
employed several relations, like \eqref{curvature} and
\eqref{fmunuminus}. A fundamental mean to eliminate redundant
monomials in $\mb$ is the use of the equations of motion  (EOM)  
satisfied by mesons and baryons 
at lowest chiral order,  ${\cal O}(q^2)$ and ${\cal O}(q)$, respectively.
The lowest order EOM satisfied by the pseudoscalar mesons 
is \cite{gl},   
\beq
D_\mu u^\mu = \frac{i}{2}\widetilde{\chi}_-~,
\label{EOM_GB}
\eeq
where $\widetilde{\chi}_-=\chi_- -\frac{1}{3}\langle \chi_-\rangle$~.
In the following, we consider $\chi_-$ as an independent structure.
The lowest order EOM satisfied  by  the  baryon matrix field is 
\bea
&&i\gamma^\mu D_\mu B-M_0 B+\frac{F}{2}\gamma^\mu 
\gamma_5[u_\mu,B]+\frac{D}{2}\gamma^\mu
\gamma_5\left(\{u_\mu,B\}-\frac{1}{3}\la\{u_\mu,B\}\ra\right)
= 0 \, , 
\label{eomb}
\eea
so that, $i\gamma^\mu D_\mu B-M_0 B= {\cal O}(q)$, as already
reported in \eqref{BChdim}. The constants $D$ and $F$ are the axial-vector 
couplings.\\ 
Another important relation for reducing   the 
${\cal O}(q^3)$  Lagrangian  is 
\beq
D^2 u_\mu=\frac{1}{4}[u_\beta,u_\mu]u^\beta-
\frac{i}{2} f^+_{\beta\mu}u^\beta+
D^\beta f_{\beta\mu}^-+\frac{i}{2}D_\mu \widetilde{\chi}_-~.
\label{d2u}
\eeq
This equation  is readily obtained by taking the derivative  of  
(\ref{fmunuminus}), using  (\ref{curvature}) and  finally applying 
the pseudoscalar meson EOM  (\ref{EOM_GB}). We will therefore 
not consider $D^2 u_\mu$   as an independent structure.

 We have  also employed  $SU(3)$ Cayley-Hamilton relations
  for reducing the number of independent 
monomials  keeping the maximum number of terms with one flavour trace.

Equations (\ref{curvature}) and (\ref{eomb}) 
 allow to derive a set of relations 
 containing different Clifford algebra elements and different number of
covariant derivatives  acting on the $B$ matrix field, namely, 
\begin{align}
&\la \bar{B}(A_1,\cdots,(A_n,\Gamma^{[\alpha]\beta}
D_\beta D^mB)\cdots)\ra \widetilde \Theta\simeq 0~,
\label{first_rel} \\
&\la \bar{B}(A_1,\cdots,(A_n,D^\alpha D^mB)\cdots) \ra \widetilde 
\Theta\simeq 
-i M_0\la\bar{B}(A_1,\cdots,(A_n,\gamma^\alpha D^mB)\cdots)
\ra \widetilde \Theta~, \label{second_rel}\\
&\la \bar{B}(A_1,\cdots,(A_n,\gamma_5 D^\alpha D^mB)\cdots)
\ra \widetilde \Theta\simeq 0~,
\label{third_rel}\\
&\la \bar{B}(A_1,\cdots,(A_n,\Gamma \gamma^\alpha 
D^\beta D^mB)\cdots)\ra \widetilde \Theta\simeq 
\la\bar{B}(A_1,\cdots,(A_n,\Gamma \gamma^\beta D^\alpha D^mB)\cdots)
\ra \widetilde \Theta~,
\label{fourth_rel} \\
&\la \bar{B}(A_1,\cdots,(A_n,\Gamma\sigma^{\alpha\beta} 
D^\lambda D^mB)\cdots)\ra \widetilde \Theta
+\la \bar{B}(A_1,\cdots,(A_n,\Gamma \sigma^{\beta\lambda} D^\alpha 
D^mB)\cdots)\ra \widetilde \Theta \nn \\
&+\la \bar{B}(A_1,\cdots,(A_n, \Gamma\sigma^{\lambda\alpha} D^\beta 
D^mB)\cdots)\ra \widetilde \Theta \simeq 0~,
\label{fifth_rel}\\
&\varepsilon_{\alpha\beta\tau\rho} \left[
\la \bar{B}(A_1,\cdots,(A_n,\Gamma\sigma^{\alpha\beta} D^\lambda 
D^mB)\cdots)\ra \right. \nn \\
&\left. 
+2 \la \bar{B}(A_1,\cdots,(A_n,\Gamma\sigma^{\beta\lambda} 
D^\alpha D^mB)\cdots)\ra\right] \widetilde \Theta \simeq 0~,
\label{sixth_rel}\\
&\varepsilon_{\alpha\beta\tau\rho}\la\bar{B}(A_1,\cdots,
(A_n,\Gamma  \sigma^{\alpha\beta} D^\tau D^mB)\cdots)\ra 
\widetilde \Theta\simeq 0~
\label{seventh_rel}
\end{align}
which will be extensively  used to reduce
the number of covariant derivatives acting on $B$. 
 Here, $\Gamma^{[\alpha]\beta}$ stands for a Clifford algebra 
basis element with either two Lorentz indices $\alpha\beta$ or one index
 $\beta$.  In these equations  
we have explicitly  shown the elements 
of the Clifford algebra basis that appear and 
$\Gamma$ is either $\ut$ or $\gamma_5$.
 The symbol $\widetilde \Theta$ refers to products of metric tensors 
 and Levi-Civita pseudotensors, while
``$\simeq$'' means  equal up to terms of higher order 
or up to terms of the same order but with less covariant
 derivatives acting on the matrix  field $B$.  
 This definition of $\simeq$ is sensible
since  those structures  of the same order but with a lower number 
of covariant derivatives are already taken into account
according to the procedure  we 
follow for writing down the list of allowed monomials.

Relations analogous to (\ref{first_rel})-(\ref{seventh_rel})
can also be obtained  for the case with two  flavour traces, because 
 what matters in their derivation is the Dirac algebra
  and the action on $B$  of covariant derivatives.
 Relations (\ref{sixth_rel})  and (\ref{seventh_rel})
are obtained from (\ref{fifth_rel})  after contracting it with the 
pseudotensor $\varepsilon_{\alpha\beta\tau\rho}$. Another
interesting result that follows 
from (\ref{first_rel}) and (\ref{second_rel})
 is that terms containing $D_\mu D^\mu D^mB$ can be discarded.
 
 Further reduction of  monomials is reached by performing more
specific manipulations --see appendix \ref{elimination} 
for  details. We finally arrive to a minimal set  
of linearly independent terms 
to ${\cal O}(q^3)$ which we present in the next section.


\section{The Effective Lorentz invariant Chiral Meson-Baryon Lagrangians
 to Order $q^3$}
\label{lagrangians}

\subsection{The Order $q^2$  Lorentz Invariant Effective Chiral Meson-Baryon Lagrangian}
 \label{SU3q2}
Following the procedure detailed in the previous sections, 
we write down the relativistic effective meson-baryon chiral Lagrangian
with 
three flavours at $\od$, 
\begin{equation}\begin{split}
\mb^{(2)}&=b_D \la\bar{B} \{ \chi_+,B\}\ra +
b_F \la\bar{B}[ \chi_+,B ]\ra +
b_0 \la\bar{B} B\ra\la\chi_+\ra +\\
&\quad
b_1 \la\bar{B} [ u^\mu,[u_\mu,B]]\ra +
b_2 \la\bar{B}\{ u^\mu,\{u_\mu,B\}\}\ra +\\
&\quad
b_3 \la\bar{B}\{ u^\mu,[u_\mu,B]\}\ra +
b_4 \la\bar{B}B\ra\la u^\mu u_\mu\ra +\\
&\quad
i b_5 \left(\la\bar{B}[u^\mu,[u^\nu,\gamma_\mu
{D}_\nu B]]\ra -
\la \bar{B}\overleftarrow{D}_\nu[u^\nu,[u^\mu,\gamma_\mu B]]\ra\right)+\\
&\quad
i b_6
\left(\la\bar{B}[u^\mu,\{u^\nu,\gamma_\mu{D}_\nu B]]\ra -
\la \bar{B}\overleftarrow{D}_\nu\{u^\nu,[u^\mu,\gamma_\mu,B]\}\ra\right)+\\
&\quad
i b_7 \left(\la\bar{B}\{u^\mu,\{u^\nu,\gamma_\mu{D}_\nu B\}\}\ra -
\la \bar{B}\overleftarrow{D}_\nu \{u^\nu,\{u^\mu,\gamma_\mu B\}\}\ra\right)+\\
&\quad
i b_8\left(\la\bar{B}\gamma_\mu{D}_\nu B\ra-
\la \bar{B}\overleftarrow{D}_\nu\gamma_\mu B\ra\right)\la u^\mu u^\nu\ra +
i d_1 \la\bar{B}\{[u^\mu,u^\nu],\sigma_{\mu\nu} B\}\ra +\\
&\quad
i d_2 \la\bar{B}[[u^\mu,u^\nu],\sigma_{\mu\nu} B]\ra +
i d_3 \la\bar{B} u^\mu\ra \la u^\nu \sigma_{\mu\nu} B\ra +
 d_4 \la\bar{B}\{ f_+^{\mu\nu},\sigma_{\mu\nu} B\}\ra +\\
&\quad
 d_5 \la\bar{B}[ f_+^{\mu\nu},\sigma_{\mu\nu} B]\ra~.
\label{lagMB2}
\end{split}\end{equation}

We compared this Lagrangian with that of
ref.~\cite{krause}. We found that 3 of the structures given in
that paper\footnote{Every structure usually 
 involves several monomials in the reduced notation of
 ref.\cite{krause}.} are redundant 
 and can be expressed in terms of the others using Cayley-Hamilton
 equation and the relation 
 (\ref{second_rel}). In ref.~\cite{frinkmas} part of the $\od$
 Lagrangian is given, the one interesting 
 for the authors' investigation, but 
 the term with coefficient $b_9$ is also redundant and using
Cayley-Hamilton equation can be written in 
 terms of the monomials proportional to $b_5-b_8$ in eq.(\ref{lagMB2}) 
or in ref.~\cite{frinkmas}.

The $SU(2)$ version of ${\cal L}^{(2)}_{MB}$ is obtained 
reducing  $\Phi$ in (\ref{mesonoctet}) to the $2\times 2$ matrix 
containing just pion fields and the matrix field $B$ 
in (\ref{baryonoctet}) to a column vector $\Psi$ collecting 
the proton   and  the neutron fields.\footnote{Of course, we 
have now to employ Cayley-Hamilton relations for $2\times 2$ matrices.}
 The external  matrix fields $s(x)$, $p(x)$, $v_\mu(x)$ and $a_\mu(x)$   
introduced in section \ref{sec:2} are also reduced to  
hermitian $2\times 2$ traceless matrices. 
In particular electromagnetic interactions are introduced through 
the external vector field $v_\mu=|e|Q A_\mu$,
 where  $Q={\rm diag}(2,-1)/3$ is the quark
electrical charge matrix and $A_\mu$ the photon
 field. Notice that in this case $\la v_\mu\ra \neq 0$
and  flavour traces of $f_{\mu\nu}^{+}$  can appear in the 
$SU(2)$ Lagrangian.  We fully agree with the $\od$ relativistic
$SU(2)$ 
meson-baryon Lagrangian
 given  in \cite{fettes98}.

\subsection{The Order $q^3$  Effective Chiral Meson-Baryon Lagrangian}
\label{SU3q3}
The meson-baryon $SU(3)$ chiral Lagrangian  at ${\cal O}(q^3)$ contains 
 84 terms that  can be generally written as
\begin{equation} 
\mb^{(3)}
= \quad \sum_{i=1}^{84} h_i \, {O}_i~.
\label{redlagMB3} 
\end{equation}
The monomials ${O}_i$
are shown in table \ref{monomials}, where we also display the vertex 
 with the lowest number of particles to which 
each interaction term gives contribution.

\renewcommand{\arraystretch}{1.5}
\begin{longtable}[r]{|c|c|c|}
\hline
$i$&$O_i$ & Contributes to vertex\\
\hline
\hline
\endhead
\hline
\caption[]{\rule{0cm}{2em}}
\endfoot
\hline
\caption[]
{\label{monomials} Minimal set of linearly independent  monomials
of the  $SU(3)$ chiral meson-baryon Lagrangian of $\ot$. On the third
column we give  the vertex  with the minimal number of mesons and photons
to which each term contributes.}
\endlastfoot
1&$i\left(\la\bar{B}\gamma_\mu
D_{\nu\rho}B[u^\mu,h^{\nu\rho}]\ra + \la\bar{B}\overleftarrow{D}_{\nu\rho}
\gamma_\mu B[u^\mu,h^{\nu\rho}]\ra\right)$& $M_1 B_1 \to M_2 B_2$\\ 
2&$i\left(\la\bar{B}[u^\mu,h^{\nu\rho}]\gamma_\mu
D_{\nu\rho}B\ra + \la\bar{B}\overleftarrow{D}_{\nu\rho}
[u^\mu,h^{\nu\rho}]\gamma_\mu B\ra\right)$ & $M_1 B_1 \to M_2 B_2$\\ 
3&$i\left(\la\bar{B}u^\mu\ra\la h^{\nu\rho} \gamma_\mu
    D_{\nu\rho}B\ra - \la\bar{B}\overleftarrow{D}_{\nu\rho} h^{\nu\rho}\ra
\la u^\mu\gamma_\mu B\ra\right)$& $M_1 B_1 \to M_2 B_2$\\
4&$i\la\bar{B}[u_\mu,h^{\mu\nu}]\gamma_\nu B\ra$& $M_1 B_1 \to M_2 B_2$\\
5&$i\la\bar{B}\gamma_\nu B[u_\mu,h^{\mu\nu}]\ra$& $M_1 B_1 \to M_2B_2$\\
6&$i\left(\la\bar{B}u_\mu\ra\la h^{\mu\nu}\gamma_\nu B\ra
-\la\bar{B}h^{\mu\nu}\ra\la u_\mu\gamma_\nu B\ra\right)$& $M_1 B_1 \to M_2B_2$\\
7&$i\la\bar{B}\sigma_{\mu\nu} D_\rho
    B\{u^\mu,h^{\nu\rho}\}\ra-i\la\bar{B}\overleftarrow{D}_\rho\sigma_{\mu\nu} 
B \{u^\mu,h^{\nu\rho}\}\ra$& $M_1 B_1 \to M_2B_2$\\
8&$i\la\bar{B}\{u^\mu,h^{\nu\rho}\}\sigma_{\mu\nu} D_\rho
    B\ra-i\la\bar{B}\overleftarrow{D}_\rho\{u^\mu,h^{\nu\rho}\}
\sigma_{\mu\nu}B\ra$& $M_1 B_1 \to M_2B_2$\\
9&$i\la\bar{B}u^\mu \sigma_{\mu\nu} D_\rho B h^{\nu\rho}\ra
-i\la\bar{B}\overleftarrow{D}_\rho u^\mu \sigma_{\mu\nu} B h^{\nu\rho}\ra$
& $M_1 B_1 \to M_2B_2$\\
10&$i\la\bar{B}h^{\nu\rho}\sigma_{\mu\nu} D_\rho B u^\mu\ra
-i\la\bar{B}\overleftarrow{D}_\rho h^{\nu\rho}\sigma_{\mu\nu}B u^\mu\ra$
& $M_1 B_1 \to M_2B_2$\\
11&$i\left(\la\bar{B}\sigma_{\mu\nu} D_\rho B\ra
-\la\bar{B}\overleftarrow{D}_\rho\sigma_{\mu\nu}B\ra\right)
\la u^\mu h^{\nu\rho}\ra$& $M_1 B_1 \to M_2B_2$\\
12&$\la\bar{B}\gamma_5\gamma_\nu B\{u_\mu u^\mu,u^\nu\}\ra$&$M_1 B_1
\to M_2 M_3 B_2$\\
13&$\la\bar{B}\gamma_5\gamma_\nu B u_\mu u^\nu u^\mu\ra$&$M_1 B_1
\to M_2 M_3 B_2$\\
14&$\la\bar{B}u_\mu\gamma_5\gamma_\nu B\{u^\mu,u^\nu\}\ra$&$M_1 B_1
\to M_2 M_3 B_2$\\
15&$\la\bar{B}u_\mu u^\mu\gamma_5\gamma_\nu B u^\nu\ra$&$M_1 B_1
\to M_2 M_3 B_2$\\
16&$\la\bar{B}\{u_\mu u^\mu,u^\nu\}\gamma_5\gamma_\nu B\ra$&$M_1 B_1
\to M_2 M_3 B_2$\\
17&$\la\bar{B}\{u^\mu,u^\nu\}\gamma_5\gamma_\nu B u_\mu\ra$&$M_1 B_1
\to M_2 M_3 B_2$\\
18&$\la\bar{B}u_\mu u^\nu u^\mu\gamma_5\gamma_\nu B\ra$&$M_1 B_1
\to M_2 M_3 B_2$\\
19&$\la\bar{B}u^\nu\gamma_5\gamma_\nu B u_\mu u^\mu\ra$&$M_1 B_1
\to M_2 M_3 B_2$\\
20&$\la\bar{B}\{u^\nu,\gamma_5\gamma_\nu B\}\ra\la u_\mu u^\mu\ra$&$M_1 B_1
\to M_2 M_3 B_2$\\
21&$\la\bar{B}[u^\nu,\gamma_5\gamma_\nu B]\ra\la u_\mu u^\mu\ra$&$M_1 B_1
\to M_2 M_3 B_2$\\
22&$\la\bar{B}\{u_\mu,\gamma_5\gamma_\nu B\}\ra\la u^\mu u^\nu\ra$&$M_1 B_1
\to M_2 M_3 B_2$\\
23& $\la\bar{B}[u_\mu,\gamma_5\gamma_\nu B]\ra\la u^\mu u^\nu\ra$&$M_1 B_1
\to M_2 M_3 B_2$\\
24&$\la\bar{B}\gamma_5\gamma_\nu B\ra\la u_\mu u^\mu u^\nu\ra$&$M_1 B_1
\to M_2 M_3 B_2$\\
25&$\la\bar{B}u_\mu\ra\la[u^\mu,u^\nu]\gamma_5\gamma_\nu B\ra
-\la\bar{B}[u^\mu,u^\nu]\ra\la u_\mu\gamma_5\gamma_\nu B\ra$&$M_1 B_1
\to M_2 M_3 B_2$\\
26&$i\la\bar{B}\gamma^\tau
  B\{[u^\mu,u^\nu],u^\rho\}\ra\ve_{\mu\nu\rho\tau}$&$M_1 B_1
\to M_2 M_3 B_2$\\
27&$i\la\bar{B}\{[u^\mu,u^\nu],u^\rho\}\gamma^\tau
  B\ra \ve_{\mu\nu\rho\tau}$&$M_1 B_1
\to M_2 M_3 B_2$\\
28&$i\la\bar{B}[u^\mu,u^\nu]\gamma^\tau B u^\rho\ra
  \ve_{\mu\nu\rho\tau}$&$M_1 B_1
\to M_2 M_3 B_2$\\
29& $i\la\bar{B}u^\rho\gamma^\tau B[u^\mu,u^\nu]\ra
  \ve_{\mu\nu\rho\tau}$&$M_1 B_1
\to M_2 M_3 B_2$\\
30&$i\la\bar{B}\gamma^\tau
  B\ra\la[u^\mu,u^\nu]u^\rho\ra\ve_{\mu\nu\rho\tau}$&$M_1 B_1
\to M_2 M_3 B_2$\\
31&$\la\bar{B}\gamma_5\gamma_\mu D_{\nu\rho}B u^\mu u^\nu u^\rho\ra
+ \la\bar{B}\overleftarrow{D}_{\nu\rho} \gamma_5\gamma_\mu B
u^\mu u^\nu u^\rho\ra$&$M_1 B_1
\to M_2 M_3 B_2$\\
32&$\la\bar{B}u^\mu\gamma_5\gamma_\mu D_{\nu\rho}Bu^\nu u^\rho\ra
+ \la\bar{B}\overleftarrow{D}_{\nu\rho} u^\mu \gamma_5\gamma_\mu B
u^\nu u^\rho\ra$&$M_1 B_1
\to M_2 M_3 B_2$\\
33&$\la\bar{B}u^\mu u^\nu\gamma_5\gamma_\mu D_{\nu\rho}B u^\rho\ra
+\la\bar{B}\overleftarrow{D}_{\nu\rho} u^\mu u^\nu \gamma_5\gamma_\mu
B u^\rho\ra$&$M_1 B_1\to M_2 M_3 B_2$\\
34&$\la\bar{B}u^\mu u^\nu u^\rho \gamma_5\gamma_\mu D_{\nu\rho}
  B\ra + \la\bar{B}\overleftarrow{D}_{\nu\rho} u^\mu u^\nu u^\rho
 \gamma_5\gamma_\mu B\ra$&$M_1 B_1\to M_2 M_3 B_2$\\
35&$\left(\la\bar{B}\{u^\mu,\gamma_5\gamma_\mu D_{\nu\rho}B\}\ra 
+ \la\bar{B}\overleftarrow{D}_{\nu\rho}\{u^\mu,\gamma_5\gamma_\mu
B\}\ra\right)\la u^\nu u^\rho\ra $&$M_1 B_1\to M_2 M_3 B_2$\\
36&$\left(\la\bar{B}[u^\mu,\gamma_5\gamma_\mu D_{\nu\rho}B]\ra 
+ \la\bar{B}\overleftarrow{D}_{\nu\rho}[u^\mu,\gamma_5\gamma_\mu
B]\ra\right)\la u^\nu u^\rho\ra $&$M_1 B_1\to M_2 M_3 B_2$\\
37&$\left(\la\bar{B}\gamma_5\gamma_\mu D_{\nu\rho}B\ra
+ \la\bar{B}\overleftarrow{D}_{\nu\rho}\gamma_5\gamma_\mu B\ra\right)
\la u^\mu u^\nu u^\rho\ra$&$M_1 B_1\to M_2 M_3 B_2$\\
38&$i\left(\la\bar{B}u^\mu\sigma^{\lambda\tau} D_\rho
  B\{u^\nu,u^\rho\}\ra - \la\bar{B}\overleftarrow{D}_\rho u^\mu
  \sigma^{\lambda\tau} B \{u^\nu,u^\rho\}\ra\right)\ve_{\mu\nu\lambda\tau}$
&$M_1 B_1\to M_2 M_3 B_2$\\
39&$i\left(\la\bar{B}\{u^\mu,\sigma^{\lambda\tau} D_\rho B\}\ra
-\la\bar{B}\overleftarrow{D}_\rho\{u^\mu,\sigma^{\lambda\tau}B\}\ra\right)
\la u^\nu u^\rho\ra \ve_{\mu\nu\lambda\tau}$&$M_1 B_1\to M_2 M_3 B_2$\\
40&$i\left(\la\bar{B}[u^\mu,\sigma^{\lambda\tau} D_\rho B]\ra
-\la\bar{B}\overleftarrow{D}_\rho[u^\mu,\sigma^{\lambda\tau}B]\ra\right)
\la u^\nu u^\rho\ra \ve_{\mu\nu\lambda\tau}$&$M_1 B_1\to M_2 M_3 B_2$\\
41&$i\left(\la\bar{B}\sigma^{\lambda\tau} D_\rho B\ra
-\la\bar{B}\overleftarrow{D}_\rho\sigma^{\lambda\tau}B\ra\right)
\la u^\mu u^\nu u^\rho\ra \ve_{\mu\nu\lambda\tau}$&$M_1 B_1\to M_2 M_3 B_2$\\
42& $i\left(\la\bar{B}u^\mu\ra\la[u^\nu,u^\rho]\sigma^{\lambda\tau}
    D_\rho B\ra+\la\bar{B}\overleftarrow{D}_\rho[u^\nu,u^\rho]\ra\la u^\mu
    \sigma^{\lambda\tau} B\ra \right)\ve_{\mu\nu\lambda\tau}$
&$M_1 B_1\to M_2 M_3 B_2$\\
43&$i\left(\la\bar{B}u^\mu\ra\la\{u^\nu,u^\rho\}\sigma^{\lambda\tau}
    D_\rho B\ra-\la\bar{B}\overleftarrow{D}_\rho\{u^\nu,u^\rho\}\ra\la u^\mu
    \sigma^{\lambda\tau} B\ra \right)\ve_{\mu\nu\lambda\tau}$
&$M_1 B_1\to M_2 M_3 B_2$\\
44&$\la\bar{B}u^\mu \gamma_5\gamma_\mu B \chi_+\ra$& $B_1 \to M_1 B_2$\\
45&$\la\bar{B}\chi_+ \gamma_5\gamma_\mu B u^\mu\ra$& $B_1 \to M_1 B_2$\\
46&$\la\bar{B}u^\mu \gamma_5\gamma_\mu B\ra\la\chi_+\ra$& $B_1 \to M_1 B_2$\\
47&$\la\bar{B}\gamma_5\gamma_\mu B u^\mu\ra\la\chi_+\ra$& $B_1 \to M_1 B_2$\\
48&$\la\bar{B}\gamma_5\gamma_\mu B\ra\la u^\mu \chi_+\ra$& $B_1 \to M_1 B_2$\\
49&$\la\bar{B}\gamma_5\gamma_\mu B\{u^\mu,\chi_+\}\ra$& $B_1 \to M_1 B_2$\\
50&$\la\bar{B}\{u^\mu,\chi_+\}\gamma_5\gamma_\mu B\ra$& $B_1 \to M_1 B_2$\\
51&$\la\bar{B}\{\chi_-,\gamma_5 B\}\ra$& $B_1 \to M_1 B_2$\\
52&$\la\bar{B}[\chi_-,\gamma_5 B]\ra$ &$B_1 \to M_1 B_2$\\
53&$\la\bar{B}\gamma_5 B\ra\la\chi_-\ra$&$B_1 \to M_1 B_2$\\
54&$\la\bar{B}\gamma_\mu B[\chi_-,u^\mu]\ra$&$B_1 M_1\to M_2 B_2$\\
55&$\la\bar{B}[\chi_-,u^\mu]\gamma_\mu B\ra$&$B_1 M_1\to M_2 B_2$\\
56&$\la\bar{B}u^\mu\ra\la\chi_-\gamma_\mu B\ra  
  -\la\bar{B}\chi_-\ra\la u^\mu\gamma_\mu B\ra$&$B_1 M_1\to M_2 B_2$\\
57&$\la\bar{B}\{D_\mu f_+^{\mu\nu},\gamma_\nu B\}\ra$ & $B_1 \to \gamma B_2$\\
58&$\la\bar{B}[D_\mu f_+^{\mu\nu},\gamma_\nu B]\ra$& $B_1 \to \gamma B_2$\\
59&$i\la\bar{B}\gamma_5\gamma_\nu B[u_\mu,f_+^{\mu\nu}]\ra$&$\gamma
B_1 \to M_2 B_2$\\ 
60&$i\la\bar{B}[u_\mu,f_+^{\mu\nu}]\gamma_5\gamma_\nu B\ra$&$\gamma
B_1 \to M_2 B_2$\\
61&$i\left(\la\bar{B}u_\mu\ra\la f_+^{\mu\nu}\gamma_5\gamma_\nu
    B\ra - \la\bar{B}f_+^{\mu\nu}\ra\la u_\mu \gamma_5\gamma_\nu
    B\ra\right)$ &$\gamma B_1 \to M_2 B_2$\\ 
62&$\la\bar{B}\gamma^\tau
  B\{u^\mu,f_+^{\nu\rho}\}\ra\ve_{\mu\nu\rho\tau}$&$\gamma B_1 \to M_2 B_2$\\ 
63&$\la\bar{B}\{u^\mu,f_+^{\nu\rho}\}\gamma^\tau
  B\ra\ve_{\mu\nu\rho\tau}$&$\gamma B_1 \to M_2 B_2$\\ 
64&$\la\bar{B} u^\mu \gamma^\tau B
  f_+^{\nu\rho}\ra\ve_{\mu\nu\rho\tau}$ &$\gamma B_1 \to M_2 B_2$\\
65&$\la\bar{B}f_+^{\nu\rho}\gamma^\tau B u^\mu\ra
  \ve_{\mu\nu\rho\tau}$ &$\gamma B_1 \to M_2 B_2$\\
66&$\la\bar{B}\gamma^\tau B\ra\la  u^\mu
  f_+^{\nu\rho}\ra\ve_{\mu\nu\rho\tau}$&$\gamma B_1 \to M_2 B_2$\\
67&$\left(\la\bar{B}[u^\mu,f_+^{\nu\rho}]\sigma^{\lambda\tau}D_\mu
    B\ra - \la\bar{B}\overleftarrow{D}_\mu [u^\mu,f_+^{\nu\rho}]
\sigma^{\lambda\tau} B\ra\right)\ve_{\nu\rho\lambda\tau}$&$\gamma B_1 \to M_2 B_2$\\
68&$\left(\la\bar{B}\sigma^{\lambda\tau}D_\mu B [u^\mu,f_+^{\nu\rho}]\ra
- \la\bar{B}\overleftarrow{D}_\mu\sigma^{\lambda\tau} B
[u^\mu,f_+^{\nu\rho}]\ra\right)\ve_{\nu\rho\lambda\tau}$&$\gamma B_1 \to M_2 B_2$\\
69&$\left(\la\bar{B}u^\mu\ra\la f_+^{\nu\rho}\sigma^{\lambda\tau}D_\mu
    B\ra + \la\bar{B}\overleftarrow{D}_\mu f_+^{\nu\rho}\ra \la
    u^\mu\sigma^{\lambda\tau} B\ra\right)
  \ve_{\nu\rho\lambda\tau}$&$\gamma B_1 \to M_2 B_2$\\
70&$\la\bar{B}\{D_\mu f_-^{\mu\nu},\gamma_5\gamma_\nu B\}\ra$&$\gamma
B_1 \to M_2 B_2$\\ 
71&$\la\bar{B}[D_\mu f_-^{\mu\nu},\gamma_5\gamma_\nu B]\ra$&$\gamma
B_1 \to M_2 B_2$\\ 
72& $\la\bar{B}\gamma_5\gamma^\tau
  B\{u^\mu,f_-^{\nu\rho}\}\ra\ve_{\mu\nu\rho\tau}$&$\gamma B_1\to M_2
  M_3 B_2$\\
73&$\la\bar{B}\{u^\mu,f_-^{\nu\rho}\}\gamma_5\gamma^\tau
  B\ra\ve_{\mu\nu\rho\tau}$&$\gamma B_1\to M_2
  M_3 B_2$\\
74&$\la\bar{B}f_-^{\nu\rho}\gamma_5\gamma^\tau B u^\mu\ra
  \ve_{\mu\nu\rho\tau}$&$\gamma B_1\to M_2 M_3 B_2$\\
75&$\la\bar{B} u^\mu \gamma_5\gamma^\tau B
  f_-^{\nu\rho}\ra\ve_{\mu\nu\rho\tau}$&$\gamma B_1\to M_2 M_3 B_2$\\
76&$\la\bar{B}\gamma_5\gamma^\tau B\ra\la  u^\mu
  f_-^{\nu\rho}\ra\ve_{\mu\nu\rho\tau}$&$\gamma B_1\to M_2 M_3 B_2$\\
77&$i\la\bar{B}[u_\mu,f_-^{\mu\nu}]\gamma_\nu B\ra$&$\gamma B_1\to M_2
M_3 B_2$\\ 
78&$i\la\bar{B}\gamma_\nu B[u_\mu,f_-^{\mu\nu}]\ra$&$\gamma B_1\to M_2
M_3 B_2$\\ 
79&$i\left(\la\bar{B}u_\mu\ra\la f_-^{\mu\nu}\gamma_\nu B\ra
    -\la\bar{B}f_-^{\mu\nu}\ra\la u_\mu\gamma_\nu B\ra\right)$&$\gamma B_1\to M_2
M_3 B_2$\\ 
80&$i\left(\la\bar{B}\sigma_{\nu\rho} D_\mu B\{u^\mu,f_-^{\nu\rho}\}\ra
-\la\bar{B}\overleftarrow{D}_\mu\sigma_{\nu\rho}B\{u^\mu,f_-^{\nu\rho}\}\ra\right)$
&$\gamma B_1\to M_2 M_3 B_2$\\
81&$i\left(\la\bar{B}\{u^\mu,f_-^{\nu\rho}\}\sigma_{\nu\rho} D_\mu B\ra
-\la\bar{B}\overleftarrow{D}_\mu\{u^\mu,f_-^{\nu\rho}\}\sigma_{\nu\rho}B\ra\right)$
&$\gamma B_1\to M_2 M_3 B_2$\\
82& $i\left(\la\bar{B}u^\mu \sigma_{\nu\rho} D_\mu B f_-^{\nu\rho}\ra
-\la\bar{B}\overleftarrow{D}_\mu u^\mu \sigma_{\nu\rho} B
f_-^{\nu\rho}\ra\right)$
&$\gamma B_1\to M_2 M_3 B_2$\\
83&$i\left(\la\bar{B}f_-^{\nu\rho} \sigma_{\nu\rho} D_\mu B u^\mu \ra
-\la\bar{B}\overleftarrow{D}_\mu f_-^{\nu\rho}\sigma_{\nu\rho} B
u^\mu \ra\right)$ &$\gamma B_1\to M_2 M_3 B_2$\\
84&$i\left(\la\bar{B}\sigma_{\nu\rho} D_\mu B\ra - 
\la\bar{B}\overleftarrow{D}_\mu\sigma_{\nu\rho}B\ra\right)
\la u^\mu f_-^{\nu\rho}\ra $&$\gamma B_1\to M_2 M_3 B_2$\\
\hline
\end{longtable}

The list of $SU(3)$ $\ot$ monomials presented in Krause's work
\cite{krause} is neither complete nor
 minimal. We have
  checked that 22 out of the 60 structures
 given in this reference   can be expressed as 
 linear combination of those already given. 
This can be done  by applying the meson EOM 
(\ref{EOM_GB}), Cayley-Hamilton equations and the relations 
(\ref{first_rel})-(\ref{seventh_rel}). 
In addition, several monomials in table \ref{monomials}
 are lacking in \cite{krause}, namely, the ones from 
 $O_7$ to $O_{10}$ and from $O_{38}$ to $O_{41}$. 

We would like to point out that the monomial $O_{41}$ , 
being of $\ot$  in the covariant counting
of (\ref{BChdim}) and (\ref{chdim_clif}), actually  starts contributing 
at ${\cal O}(q^4)$   to meson-baryon amplitudes 
in a non-covariant  chiral counting.  To see this,  notice
that  in a non-covariant  counting, the ${\cal O}(q^3)$ contributions from
$O_{41}$  are generated  
when the  index $\rho$ is temporal and $\lambda$ and $\tau$ are both
spatial. Then in this case one has 
\bea
i\left(\la\bar{B}\sigma^{ij} D_0 B\ra
-\la\bar{B}\overleftarrow{D}_0\sigma^{ij}B\ra\right)
\la u^\mu u^\nu u^0 \ra \ve_{\mu\nu i j } = 0.
\eea

 We have also derived the $SU(2)$ version of
the $\mb^{(3)}$ meson-baryon Lagrangian in the same way as  we did  
for the $\od$ Lagrangian (\ref{lagMB2}) and
found a full agreement with the one obtained in \cite{fettes98}.
 
\section{Summary and Conclusions}
\label{conc}

As already mentioned in the Introduction, in the literature can be found 
 several one loop calculations performed in baryon CHPT employing 
parts of the $\ot$
three flavour Lagrangian (\ref{redlagMB3}). However, in this work, we 
derived for the first time the complete
$\od$ and $\ot$ Lorentz invariant
 $SU(3)$ effective meson-baryon chiral 
Lagrangians, eqs.~\eqref{lagMB2} and \eqref{redlagMB3}, respectively. 
We both  reduced the
number of independent monomials given in previous studies 
\cite{krause,frinkmas} and identified missing terms \cite{krause}.
There is  
perfect agreement between the $SU(2)$ reduction of the $\od$ 
and $\ot$ relativistic Lagrangians we obtained
and those of ref.\cite{fettes98}. We also gave ${\cal L}^{(2)}_{MB}$ and 
${\cal L}^{(3)}_{MB}$ in a way that it is exactly invariant under 
charge conjugation.

\section*{Acknowledgements}

 This work has been supported in part by the MEC (Spain) and FEDER (EC) Grants
  Nos. FPA2003-09298-C02-01 (J.P.), FPA2004-03470 (J.A.O. and M.V.),  the 
  Fundaci\'on  S\'eneca grant Ref. 02975/PI/05  (J.A.O. and M.V.), the European Commission
(EC) RTN Network EURIDICE under Contract No. HPRN-CT2002-00311 and the HadronPhysics I3
Project (EC)  Contract No RII3-CT-2004-506078 (J.A.O.) and by  Junta de Andaluc\'{\i}a Grants Nos.
FQM-101 (J.P. and M.V.) and FQM-347 (J.P.). M.V. also acknowledges financial support from the 
Fundaci\'on S\'eneca (Murcia) and the Departamento de F\'{\i}sica  Te\'orica y del Cosmos,
Universidad de Granada, for the warm hospitality.

\section*{Appendices}

\appendix

\section{Elimination of Monomials}
\label{elimination}
In this appendix we show details on how
we have further reduced the number of monomials
by applying the  
relations  (\ref{curvature}) and (\ref{fmunuminus}) from  right to left 
and  then reintroducing covariant derivatives.
Integrating  by parts and neglecting  total derivatives, 
one then applies the baryon EOM (\ref{eomb})
 and its hermitic conjugate, and checks  whether such monomials 
are independent  or a combination of other ones already considered. 
We have also employed  (\ref{fifth_rel}) with $\Gamma=\gamma_5$
as explained below.

In this way, by applying eq.~\eqref{curvature}, 
we can  remove the following monomial
\beq
i\la\bar{B}\{[u_\sigma,[u_\rho,u_\eta]],\sigma_{\alpha\beta}D^\sigma B\}
\ra\varepsilon^{\al\be\rh\et}~.
\eeq
 As an intermediate step in the elimination of this monomial,
 we used the identity
 \bea
\sigma_{\al\be}\varepsilon^{\al\be\rh\et}&=&2i\gamma_5\sigma^{\rh\et}\nn\\
&=&2\ga_5(g^{\rh\et}-\ga^\rh\ga^\et).
\label{sigmarel}
 \eea
This is employed in order to contract  
space-time indices of covariant derivatives 
acting on $B$ or $\bar{B}$  with those of  
$g^{\rh\et}=(\ga^\rh\ga^\et+\ga^\et\ga^\rh)/2$ and of $\ga^\rh\ga^\et$ 
in the second line of (\ref{sigmarel}) 
and then apply the baryon EOM \eqref{eomb}.

Employing the first line of (\ref{sigmarel}), 
together with the cyclic relation (\ref{fifth_rel}) 
with $\Gamma=\ga_5$, we can relate the two  monomials,
\bea
&&\varepsilon_{\al\be\si\rh}\la\bar{B}\{[f_+^{\si\rh},u^\nu],\si^{\al\be}
D_\nu B\}\ra~,\nn\\
&&\varepsilon_{\al\be\si\rh}\la\bar{B}\{[f_+^{\si\nu},u^\rho],\si^{\al\be}
D_\nu B\}\ra~
\label{a.3}
\eea
and express the latter in terms of the former, modulo terms of 
higher order or  terms   already considered  with less covariant derivatives 
acting on $B$. \\
One can proceed in a similar way for the monomials 
involving two flavour traces,
\bea
&&\varepsilon_{\al\be\si\rh}\left(\la \bar{B}\si^{\al\be}u_\et\ra\la 
f_+^{\si\rh}D^\et B\ra-
\la\bar{B} f_+^{\si\rh}\ra\la u_\et\si^{\al\be} 
D^\et B\ra\right)~,\nn\\
&&\varepsilon_{\al\be\si\rh}\left(\la \bar{B}\si^{\al\be}u_\et\ra\la f_+^{\et\si}
D^\rh B\ra-\la\bar{B} f_+^{\et\si}\ra\la u_\et\si^{\al\be}
 D^\rh B\ra\right)~
\label{eq1}\eea
and remove the second  monomial in \eqref{eq1}.\\
The elimination of 
\bea
&&\varepsilon_{\al\be\si\rh}\la\bar{B}\{D^\nu f_-^{\si\rh},\si^{\al\be}
D_\nu B\}\ra~,\nn\\
&&{\rm and} \,\, 
\varepsilon_{\al\be\si\rh}\la\bar{B}\{D^\rh f_-^{\nu\si},\si^{\al\be}
D_\nu B\}\ra~,
\label{eq2}
\eea
is done in two steps.  First, we 
 write down the second monomial above in terms of the first one
and  others already considered by applying (\ref{sigmarel}) and the cyclic 
relation (\ref{fifth_rel}),  with $\Gamma=\gamma_5$
as done for (\ref{a.3}) and (\ref{eq1}). Next, 
the first monomial is  removed by employing from right to left
 (\ref{fmunuminus}), and then applying 
repeatedly the baryon EOM together with (\ref{curvature}) and 
(\ref{third_rel}) . 

\section{Field Transformations and Use of EOM}\label{equivalence}
In section \ref{construction} we employed baryon EOM as a mean to eliminate
redundant structures in the construction of the $\od$ and $\ot$
effective meson-baryon Lagrangians. Here we discuss the equivalence
between using EOM and performing baryon field transformations in order
to minimize the number of terms in such Lagrangians. In the mesonic sector 
this equivalence was demonstrated in refs.~\cite{EOM_FT_mes},
while within $SU(2)$ baryon CHPT this issue was addressed in
ref.~\cite{fettes98}.

Suppose that we are dealing with the list of $\od$
meson-baryon monomials,
in which appears an operator of the form
\begin{equation}
{\mathcal O} = i\left(\la\bar{B} A \sh D B\ra
- \la\bar{B} \sh \overleftarrow{D}A B \ra \right )\widetilde\Theta~,
\label{EOM_op}
\end{equation}
where $A$ is of $\od$ and can be either a single chiral field  or a
product or a (anti)commutator of chiral fields. For the sake of simplicity, we take 
$(-1)^{c_A}=(-1)^{h_A}=1$. Our goal is getting rid of the term in
eq. \eqref{EOM_op}, which contains a structure present in the baryon EOM 
\eqref{eomb} and in its hermitic conjugate. To this end, we perform
the following transformation on the baryon fields
\begin{equation}\begin{split}
B&\longrightarrow B^\prime = (1-A)B~,\\
\bar{B}&\longrightarrow \bar{B}^\prime = \bar{B}(1-A)~,
\label{FT}
\end{split}\end{equation}
which is actually a field translation.
Let us consider the effect produced by this transformation in the
${\cal O}(q)$ effective meson-baryon Lagrangian,
\begin{equation}
\mb^{(1)}=\la\bar{B}(i\gamma^\mu D_\mu-M_0)B\ra+
\frac{D}{2}\la\bar{B}\gamma_\mu\gamma_5\{u^\mu,B\}\ra
+\frac{F}{2}\la\bar{B}\gamma_\mu\gamma_5[u^\mu,B]\ra~.
\label{lagmb1}
\end{equation}
Inserting the new fields $\bar{B}^\prime,~B^\prime$, we obtain
\begin{equation}
\mb^{(1)}\longrightarrow \mb^{(1)}
- i\left(\la\bar{B} A \sh D B\ra
- \la\bar{B} \sh \overleftarrow{D}A B \ra \right )\widetilde\Theta
+ 2 M_0 \la\bar{B}A B\ra +\ot~.
\label{lagmb1_FT}
\end{equation}
The second term in the r.h.s. exactly cancels the operator in
eq.~\eqref{EOM_op}. This elimination corresponds to the relation
\eqref{first_rel} derived directly using the baryon EOM.
The same procedure carried out at $\od$ can be repeated similarly at
$\ot$ and higher. Applying then Dirac algebra manipulations 
and finally the field translation \eqref{FT}, we can obtain the
relations \eqref{first_rel}-\eqref{seventh_rel}, which allow to
eliminate monomials with covariant derivatives acting on the baryon
fields in favor of terms with less covariant derivatives.

With the field transformation \eqref{FT} we induce
changes in higher order terms. However, since in an effective
field theory we generate the list of all possible terms obeying the
required symmetries, all these modifications only shift the
values of some unknown coupling constants, but not the structure of
the corresponding monomials. 

The basic motivation for  employing field transformations to minimize
the number of terms in effective Lagrangians is the equivalence
theorem. This theorem states that in renormalized field theories
$S$-matrix elements (i.e. physical observables) are independent of the
choice of the interpolating fields or, equivalently, are invariant
under field transformations (provided the transformations satisfy certain
properties) \cite{eq_th_QFT}. The equivalence theorem was extended to
effective field theory in refs.~\cite{eq_th_EFT}.

\end{document}